\documentclass[11pt]{article}
\usepackage{arxiv}
\usepackage{amssymb,amsmath,amsthm}

\usepackage{bm}
\usepackage{subcaption}
\usepackage{multirow}
\usepackage{mathtools}
\usepackage{threeparttable}
\usepackage{booktabs}
\usepackage{blkarray, bigstrut} %
\usepackage{arydshln}
\usepackage{rotating}
\usepackage[numbers]{natbib}
\usepackage{tikz}
\usepackage{algorithm}
\usepackage{algpseudocode}%
\usepackage{xcolor} 
\usepackage{listings}
\usepackage[colorlinks,citecolor=blue]{hyperref}

\definecolor{myblue}{rgb}{0.0, 0.0, 1.0} 
\definecolor{mygreen}{rgb}{0.0, 0.5, 0.0} 
\definecolor{mypurple}{rgb}{0.5, 0.0, 0.5} 
\definecolor{mygray}{rgb}{0.8, 0.8, 0.8} 

\lstdefinestyle{customr}{
    language=R,
    backgroundcolor=\color{lightgray!10}, 
    basicstyle=\footnotesize\ttfamily, 
    breaklines=true, 
    numbers=left, 
    numberstyle=\tiny\color{gray}, 
    keywordstyle=\color{mypurple}, 
    commentstyle=\color{mygreen}, 
    stringstyle=\color{mypurple}, 
    identifierstyle=\color{mypurple}, 
}

\newcommand{\proglang}{\text}
\newcommand{\pkg}{\text}

\newcommand{\code}[1]{\textbf{$#1$}}

\begin{document}
\title{\pkg{trajmsm}: An R package for Trajectory Analysis and Causal Modeling}

\author{Awa Diop\\Université Laval \And Caroline Sirois\\Université Laval \And Jason R. Guertin\\Université Laval \AND Mireille E. Schnitzer\\Université de Montréal \And James M. Brophy \\McGill University \\ \AND  Denis Talbot\\Université Laval}
\maketitle

\begin{abstract}
The \proglang{R} package \pkg{trajmsm} provides functions designed to simplify the estimation of the parameters of a model combining latent class growth analysis (LCGA), a trajectory analysis technique, and marginal structural models (MSMs) called LCGA-MSM. LCGA summarizes similar patterns of change over time into a few distinct categories called trajectory groups, which are then included as ``treatments" in the MSM.  
MSMs are a class of causal models that correctly handle treatment-confounder feedback. The parameters of LCGA-MSMs can be consistently estimated using different estimators, such as inverse probability weighting (IPW), g-computation, and pooled longitudinal targeted maximum likelihood estimation (pooled LTMLE). These three  estimators of the parameters of LCGA-MSMs are currently implemented in our package. In the context of a time-dependent outcome, we previously proposed a combination of LCGA and history-restricted MSMs (LCGA-HRMSMs). Our package provides additional functions to estimate the parameters of such models. Version 0.1.3 of the package is currently available on CRAN.
\end{abstract}

\quad

\begin{center}
\end{center}

\clearpage

\sloppy

\section{Introduction}

In the analysis of longitudinal data with a time-varying treatment (or exposure), time-varying covariates may have the dual role of confounders and intermediaries in the association between treatment and outcome. Traditional statistical methods, such as adjusting for baseline covariates, or adjusting for both baseline and time-varying covariates in an outcome regression model are inadequate to deal with such treatment-confounder feedback  \citep{3hernan2000marginal}. The former solution leads to residual confounding, because time-varying confounders remain unadjusted, whereas the latter leads to overadjustment because time-varying covariates are on the causal pathway between previous treatments and the outcome. The marginal structural model (MSM) methodology is increasingly used to adequately handle treatment-confounder feedback \citep{hernan2010causal}. MSMs model the counterfactual outcome as a function of previous treatments and optionally baseline covariates, allowing for the comparison between various counterfactual treatment scenarios. For example, one could compare the proportion of individuals who would have developed a disease had everyone been treated for the full follow-up versus had no one been treated at any point of the follow-up.

When there are multiple time points, one challenge that arises is how to summarize the relationship between the counterfactual outcome and the various possible treatment patterns. For a binary treatment measured at $K$ different time points, there are $2^K$ possible treatment patterns. This makes it infeasible, or at least impractical, to individually compare all possible pairs of treatment patterns. To navigate this complexity, trajectory analysis methods can be used to identify a few trajectory groups that we define within an MSM for more flexibility~\citep{nagin2005group, diop2023marginal}. One of the most popular trajectory analysis techniques is latent class growth analysis (LCGA), which can be viewed as a dimension reduction technique where individual treatment trajectories are summarized into a few trajectory groups~\citep{nagin2005group, franklin2013group}. The use of clustering to define trajectory groups is an effective way to smooth over complex treatment trajectories, thereby facilitating the definition of a meaningful MSM.

In previous work, we proposed an approach to combining LCGA and MSMs  by first summarizing the individual treatment trajectories into a few groups using LCGA and then fitting an MSM conditional on the treatment trajectories, an approach we call LCGA-MSM \citep[]{diop2023marginal}. We have illustrated the ability of our approach to yield consistent estimates in the presence of treatment-confounder feedback and the validity of the inference (i.e., 95\% confidence intervals). In contrast, we have also shown that estimating the effect of the trajectory groups using traditional adjustment methods can yield highly biased estimates with respect to the target MSM coefficients. 

However, the LCGA-MSM approach models trajectories using only one time interval, with the outcome measured at the end of follow-up. This limits data usage, excluding individuals who experienced the outcome during the interval and overlooking potential changes in trajectories during the follow-up period. To address these limitations and make better use of the data, we introduced a combination of history-restricted MSMs (HRMSMs) and LCGA \citep{diop2024history}. LCGA-HRMSM can be seen as a repeated application of LCGA-MSM across multiple time intervals, allowing for the consideration of time-dependent outcomes. To estimate the parameters of both LCGA-MSMs and LCGA-HRMSMs, we proposed three different estimators: inverse probability weighting (IPW), g-computation (or g-formula), and pooled longitudinal targeted maximum likelihood estimation (pooled LTMLE).

In this paper, we present the \proglang{R} package \pkg{trajmsm} that implements the LCGA-MSM methods. The \pkg{trajmsm} package offers the possibility of estimating the parameters of an LCGA-MSM with IPW, g-computation and pooled LTMLE. For IPW, three types of outcomes are currently supported: continuous, binary and time-to-event (survival). In the case of g-computation and pooled LTMLE, only binary outcomes are currently supported. In the remaining, we present an overview of the LCGA-MSM framework before presenting the features of the \pkg{trajmsm} package. The functions we have proposed are intended to facilitate the estimation process. Version 0.1.3 of the package is currently available on CRAN. Before continuing, we note that while we use a terminology and examples that are focused on medical sciences, the methods we present can be applied to various fields where interest lies in estimating the effect of a time-varying ``treatment'' or exposure on an outcome after adjusting for time-varying confounders. Notably, LCGA has been applied in many fields other than health, such as criminology, psychology and education \citep{nagin2010group,eisenlohr2020there,suerken2016marijuana}. 

\section{Overview of the LCGA-MSM}

\begin{figure}[h!]
\centering
\resizebox{.75\linewidth}{!}{
\begin{tikzpicture}
    \node (V) at (-6, -2)  {$\bm{V}$};
	\node (L1) at (-3.5, -3)  {$\bm{L}_{t=1}$};
	\node (A1) at (-3, 0)  {$\bm{A}_{t=1}$};
	\node (L2) at (0.5,-3)  {$\bm{L}_{t=2}$};
	\node (L3) at (4.5,-3)  {$\bm{L}_{t=3}$};
	\node (A2) at (1, 0)   {$\bm{A}_{t=2}$};
	\node (A3) at (5, 0)   {$\bm{A}_{t=3}$};
	\node (Y) at (8, -1)   {$\bm{Y}$};

	\draw[color = black,line width = 2,->] (V) to (L1);
	\draw[color = black,line width = 2,->] (V) to (A1);
	\draw[color = black,line width = 2,->] (L1) to (A1);
	\draw[color = black,line width = 2, ->](L1) to (L2);
	\draw[color = black,line width = 2, ->](L2) to (L3);
	\draw[color = black,line width = 2, ->] (A1) to (A2);
	\draw[color = black,line width = 2, ->] (A2) to (A3);
	\draw[color = black,line width = 2, ->] (A1) to (L2);
	\draw[color = black,line width = 2, ->] (A2) to (L3);
	\draw[color = black,line width = 2, ->] (L2) to (A2);
	\draw[color = black,line width = 2, ->] (L3) to (A3);

	\draw[color = black,line width = 2,->] (A3) to (Y);
	\draw[color = black,line width = 2,->] (L3) to (Y);
\end{tikzpicture}}
\caption{Directed acyclic graph representing the data-structure with $A_1, A_2, A_3$ time-varying treatment, $L_1,L_2,L_3$, $V$ baseline characteristics, and $Y$ outcome. Some arrows are left out for simplicity of presentation.}
\label{fig_tp1}
\end{figure}
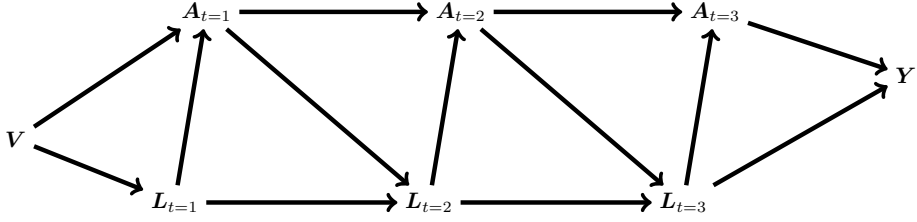

Let $\bar{A}_{t}=(A_{1},\ldots A_{t})$ be the history of observed binary treatments up to time $t = 1, \ldots, K$, $\bar{L}_{t}=(L_{1},\ldots, L_{t})$ the observed covariates up to time $t$, $V$ the baseline variables and $Y$ the observed outcome measured at the end of the follow-up period. The generic time-ordered structure is given by $(V, L_1, A_1,C_1,...,L_K, A_K,C_K,Y)$ where $C_t$ represents the missing indicator. When $C_t = 1$ censoring occurs at or before time $t$ and $C_t = 0$ means absence of censoring.  All past variables are allowed to affect all future variables. Figure \ref{fig_tp1} depicts the possible relations between these variables for a study with $K=3$ time points using a simplified directed acyclic graph (DAG). The DAG is presented as a complement to the generic data structure where some arrows and the censoring nodes are left out for simplicity of presentation. Notably, this is only one of the data structures that is permissible in our package \pkg{trajmsm}.

\subsection{First step: Creating trajectory groups with LCGA}

The first step of the LCGA-MSM approach is to use LCGA to group together individuals sharing a similar profile of their treatment pattern $\bar{A}_K$ over time, forming the trajectory groups. Let $\bar{A}_{iK}=(A_{i1},\ldots A_{iK})$ be the observed binary treatment up to time $K$ for the $i^{th}$ individual and $J$ the number of trajectory groups. Denote $z_i$ the group-membership or trajectory group for the $i^{th}$ individual.  The probability of treatment at each time in each trajectory group is modeled as a polynomial function of the follow-up time. For example, if a cubic relation was assumed for the $j^{th}$ group, we could use the regression model $\text{logit}(P(A_{it}|z_i = j)) = \theta_0^j + \theta_1^j t + \theta_2^j  t^2 + \theta_3^j  t^3$ where the parameters $\theta^{j}, j=1,\ldots, J$ summarize how the treatment probability varies over time for the $j^{th}$ group. To recover the marginal distribution of the treatment, the following model is considered:
\begin{equation}
P(\bar{A}_{iK})=\sum_{j=1}^J\pi_j \prod_{t=1}^K P(A_{it}|z_i=j).
\label{mixt_chap5}
\end{equation}
In Equation (\ref{mixt_chap5}), the $\pi_j, j=1,\ldots, J$ are the marginal probabilities that an individual belongs to the $j^{th}$ trajectory group, also called the mixture proportions. Equation (\ref{mixt_chap5}) is a mixture model where the individual treatment trajectory of each subject is represented as a mixture (i.e., a weighted average) of the different trajectory groups. In practice, when fitting an LCGA, both the groups' polynomial order and the number of groups must be specified. However, it is possible to choose both of these hyper-parameters using a data-assisted procedure, for example using the Bayesian Information Criterion (BIC) (see Section \ref{s:iden_traj}). 

Note that to avoid identifiability problems, the number of groups for a binary treatment should be chosen such that $J < (K + 1)/2$ \citep{grun2008identifiability}. For example, if $K=6$, the maximum number of groups that can be identified is $J = 3$. To assess how well a given LCGA fits the data, various statistics can be used, including average posterior probabilities, odds of correct classification and the entropy information criterion (EIC)  \citep{van2020overview}. Conditional or posterior probabilities are computed post-model fitting and are used to assign subjects into a trajectory group. The $j$th conditional probability is given by:
\begin{equation}
P(z_i=j|\bar{A}_{iK})=\dfrac{\pi_{j}P(\bar{A}_{iK}|z_i=j)}{\sum_{j^{'}=1}^J\pi_{j^{'}}P(\bar{A}_{iK}|z_i=j^{'})}. \label{eq2}
\end{equation}
To assign individuals to a group, the highest posterior probability is often used as a criterion \citep{nagin2010group}. There are other classification approaches such as modal and random assignment \citep{vermunt2010latent}.  However these approaches are not implemented in the current version of the package \pkg{trajmsm}.
 
\subsection{Second step: Expressing the causal effect as a function of the trajectory categories using MSMs}

The second step of the LCGA-MSM approach consists of estimating the effect of the trajectory category on the outcome. The causal parameter of interest $\beta$ is defined as the vector that minimizes the statistical distance between the true nonparametric MSM $m^*$ and the working model $m$ (which we define below). In other words, the parameter of interest can be thought of as the best possible approximation, using a relatively simple working model that is a function of the trajectory, of the true (unknown) MSM. Note that the nonparametric MSM is defined without making assumptions regarding the full data distribution, i.e., the distribution of the observed and counterfactual data which we will present shortly \citep{neugebauer2007nonparametric}.

It should be noted that we do not assume the existence of ``true" trajectory groups in the population. Rather, LCGA is considered a convenient dimension reduction technique, useful for reducing the numerous (possibly $2^K$) potential treatment patterns into a few trajectory groups. As also explained by \citet{nagin2005group}, the groups can be viewed as points of support for the true underlying distribution of the individual trajectories. Thus, a parametric MSM that relates the counterfactual outcome to the trajectory categories will most likely be misspecified. It is to circumvent this challenge that we define a working model $m$ that approximates the true nonparametric MSM $m^*$. By doing this, we do not assume that the specified functional relation between the outcome and the trajectory categories is correct though we hope that it offers a ``useful'' representation of $m^*$. 

More formally, let $Y^{\bar{a}_K}$ be the counterfactual outcome under the treatment pattern $\bar{a}_K$. For example, consider a binary treatment (0/1) and $K=3$. Thus, there are $2^3 = 8$ possible treatment patterns $\bar{a}_K$: $000$, $001$, $010$, $100$, $011$, $101$, $110$ and $111$ and $2^3$ corresponding counterfactual or potential outcomes $Y^{\bar{a}_K}$ for each individual, each one a random variable. Conceptually, these counterfactual outcomes represent the outcome variable had the treatment pattern been set to $\bar{a}_K$, possibly contrary to fact. The nonparametric MSM $m^*$ without assumptions on the joint distribution of all counterfactual variables $X = (\bar{L}^{\bar{a}},Y^{\bar{a}}: {\bar{a} \in \mathcal{A}})$, i.e. the full data distribution $F_X$, can be expressed as:

\begin{equation}
E_{Y^{\bar{a}_K}}(Y^{\bar{a}_K}) = m^{*}(\bar{a}_K).
\label{msmgen}
\end{equation}

To define the working MSM, it is first necessary to note that the trajectory categories are a function of the individual trajectories $\bar{a}_K$. As such, we can express the working MSM as a function of the trajectory categories. First, each possible treatment pattern $\bar{a}_K \equiv \bar{a}$ is assigned to a category based on the output of the LCGA conducted in the first step. Let $z^*(\bar{a}) = (z^*_1(\bar{a}) , ..., z^*_J(\bar{a}) )$ be dummy variables indicating to which category the treatment pattern $\bar{a}_K$ was assigned to. When the outcome is continuous, the working model can be expressed as follows: \begin{equation} m(z^*) = \beta_0 + \beta_1 z^*_1(\bar{a})  + ... + \beta_{J-1} z^*_{J-1}(\bar{a}) ,\end{equation} 
and the parameters $\beta_{j}, j = 1, \ldots, J-1$ represent the causal mean differences between each category and the reference category, $J$. When the outcome $Y$ is binary, we can consider the following working model:
\begin{equation} m(z^*) =\text{expit}(\beta_0+\beta_1z^*_1(\bar{a}) +\beta_2z^*_2(\bar{a}) +\ldots+\beta_{J-1}z^*_{J-1}(\bar{a}) ).\end{equation}
The parameters $\exp(\beta_{j}), j = 1, \ldots, J-1$ represent in this case the causal odds ratios with the $J$th category as reference category.

\subsubsection{Case of a time-to-event outcome}

When  $Y$ is a time-to-event outcome, we define the working causal model using a Cox proportional hazards MSM. The data consist of the following elements: $\{ T_i, \delta_i,\bar{L}_i, \bar{C}_i, \bar{A}_i, V_i \}, i = 1, \ldots, n$, where $T_i$ is the minimum between the observed time-to-event and outcome censoring time, $\delta_i$ is the event indicator (1 if the event occurs, 0 otherwise), $V_i$ represents the baseline covariates, $\bar{L}_i$ the observed time-varying covariates, $\bar{C}_i$ are censoring indicators during the treatment follow-up time, and $\bar{A}_i$ denotes the treatment regime for individual $i$. The nonparametric MSM $m^*(\bar{a}, t)$ represents the counterfactual hazard at time $t$ under the treatment pattern $\bar{a}$, expressed as:
\begin{equation}
m^*(\bar{a}, t) = h^{\bar{a}}(t),  
\end{equation}
or alternatively, in terms of survival probabilities:
\begin{equation}
m^*(\bar{a}, t) = S^{\bar{a}}(t) = \exp\left( -H^{\bar{a}}(t) \right),
\end{equation}
where $S^{\bar{a}}(t)$ is the counterfactual survival probability, and $H^{\bar{a}}(t)$ is the cumulative hazard function under treatment pattern $\bar{a}$. We model the counterfactual hazard using a Cox proportional hazards model, where the baseline hazard function is $h_0(t)$, and the counterfactual hazard function under treatment pattern $\bar{a}$ is given by:
\begin{equation}
m(z^*, t) \equiv h^{\bar{a}}(z^*, t) = h_0(t) \exp\left( \beta_1 z^*_1(\bar{a}) + \ldots + \beta_{J-1} z^*_{J-1}(\bar{a}) \right).
\end{equation}
Where $z^*_j(\bar{a})$ represents the covariates under treatment $\bar{a}$, and the parameters $\exp(\beta_j)$, for $j = 1, \ldots, J-1$, correspond to the causal hazard ratios relative to the reference category. This formulation allows us to estimate the relative hazard of the event (e.g., cardiovascular event or death) for individuals exposed to different treatment regimes compared to the reference category. The model $m(z^*, t)$ is then used to model the counterfactual hazard $h^{\bar{a}_K}(t)$, which describes the hazard of the event occurring at time $t$ under the counterfactual treatment pattern $\bar{a}_K$.

\subsection{Estimation of the parameters of LCGA-MSMs}\label{s:estim}

\subsubsection{Inverse probability weighting (IPW)}
To estimate the parameters of the LCGA-MSM, the first proposed estimator is inverse probability weighting (IPW). IPW is a popular estimation method that creates a pseudo-population where treatment at each time point is independent of measured confounders and treatment history \citep{cole2008constructing}. Like other approaches that estimate the parameters of MSMs, IPW is used to control for confounding bias and selection bias. The weights assigned to individuals are proportional to the inverse of their probabilities of receiving the treatment that they actually received (inverse probability of treatment weighting) or of being included in the sample (inverse probability of censoring weighting) \citep[]{cole2008constructing, hernan2010causal}. IPW also has the advantage of clearly separating the design phase from the analysis phase \citep{austin2011introduction}. 

 In \pkg{trajmsm}, unstabilized and stabilized weights are available. Note that the weights can take various forms. Unstabilized weights aim to mimic a randomized study without any losses to follow-up where all treatment patterns are equally likely. However, unstabilized weights are known to yield more variable estimates due to the presence of extreme weights \citep{cole2008constructing}. Stabilized weights also aim to mimic a randomized study without losses to follow-up, but with unequal randomization probabilities: treatment patterns that are more commonly observed in the data are given more weight than more infrequent treatment patterns. Stabilized weights are generally used to obtain narrower confidence intervals. Denote by $W$, $SW$, $WC$ and $SWC$, the unstabilized inverse probability weights for treatment, the stabilized inverse probability weights for treatment, the unstabilized inverse probability weights for censoring and the stabilized inverse probability weights for censoring, respectively. These weights have the following expressions:
$$W  =\dfrac{1}{\prod_{t=1}^K P(A_{t}|\bar{A}_{t-1},\bar{L}_{t}, C_{t-1} = 0)},
 SW  =\dfrac{\prod_{t=1}^K P(A_t|\bar{A}_{t-1}, C_{t-1} = 0)}{\prod_{t=1}^K P(A_{t}|\bar{A}_{t-1},\bar{L}_{t}, \bar{C}_{t-1} = 0)}.$$\\
$$WC  =\dfrac{1}{\prod_{t=1}^K P(C_{t}=0|\bar{A}_{t-1},\bar{L}_{t}, C_{t-1} = 0)}, SWC =\dfrac{\prod_{t=1}^K P(C_t =0|\bar{A}_{t-1}, \bar{C}_{t-1} = 0)}{\prod_{t=1}^K P(C_{t}= 0|\bar{A}_{t-1}, \bar{C}_{t-1} = 0,\bar{L}_{t})}.$$
When there is censoring, the product of $W\times WC$ is the unstabilized weight and the product $SW\times SWC$ is the stabilized weight. To estimate the parameters of LCGA-MSMs using IPW, first the trajectory groups are identified using LCGA. The reference group needs to be identified. Then, logistic regression models are fitted to estimate the probabilities involved in the IPW. Finally, a weighted regression model relating the outcome to the trajectory groups is fitted. A robust (sandwich) variance estimator is used to account for estimated weights, which yields conservative inferences \citep{lok2021estimating}.


\subsubsection{G-computation}

There is an important difference between estimation using IPW compared to using g-computation or pooled LTMLE. Indeed, implementation of g-computation and pooled LTMLE in the LCGA-MSM framework follows three major steps: estimation of the counterfactual mean under each possible treatment pattern $\bar{a}_K$, assigning each treatment pattern to a trajectory category, regression of the estimated counterfactuals means on the trajectory categories. In what follows, we present how to use g-computation and pooled LTMLE estimation approaches in an LCGA-MSM framework.

G-computation is an estimation method for the parameters of MSMs where the outcome is modeled (instead of the treatment as in IPW).  G-computation is a generalization of the classical standardization formula to a setting with a time-varying treatment and time-varying confounders \citep{hernan2010causal}. It can be seen as simulating what the outcomes would be under each treatment pattern $\bar{a}_K$. For each follow-up period and under a fixed treatment pattern, the mean outcome is estimated conditional on  covariate history \citep{hernan2010causal}. The counterfactual mean $E(Y^{\bar{a}_K})$ can be estimated using a backward updating approach with a final marginalization step \citep{van2011targeted, bang2005doubly}. For each of the $2^K$ treatment patterns, counterfactual means can be obtained through Algorithm \ref{alg:1} for a binary outcome. The estimated counterfactual mean outcome for each treatment pattern is then regressed onto the indicator variables for each non-reference trajectory category, where the treatment pattern is a member of the category indicated. Unfortunately, no closed-form variance expression is available for the general g-computation algorithm \citep{hernan2010causal}. In the \pkg{trajmsm} package, bootstrap methods are used to estimate the standard errors of the g-computation algorithm.

\begin{algorithm}
\caption{Algorithm for estimating parameters of LCGA-MSMs using g-computation}\label{alg:1}
\begin{algorithmic}
\State 1. Fit an LCGA.
\For{each treatment patterns $\bar{a}_K$}
\State  2a. Set $Q^{\bar{a}_K}_{K+1} = Y$
 \State 2b. Fit an outcome model for 
      $E(Q^{\bar{a}_K}_{K+1}|\bar{A}_K,\bar{L}_K, C_K=0) =\text{expit}(\gamma_0+\gamma_1\bar{A}_{K}+\gamma_2\bar{L}_{K}).$
  \State 2c. Compute the predicted value $\hat{Q}^{\bar{a}_K}_K=\hat{E}(Q^{\bar{a}_K}_{K+1}|\bar{A}_{K}=\bar{a}_K,\bar{L}_{K}, C_K=0)$; using the estimated $\hat{\gamma}_0, \hat{\gamma}_1, \hat{\gamma}_2$ from step 2b, then, similarly:
\For{$t = K-1$ to 1}
 \State 3a. Regress the outcome onto the time-varying treatment and covariates: $E(\hat{Q}^{\bar{a}_K}_{t+1}|\bar{A}_{t},\bar{L}_{t}, C_t=0) = \text{expit}(\gamma_0+\gamma_1\bar{A}_{t}+\gamma_2\bar{L}_{t})$
 \State 3b. Compute the predicted value $\hat{Q}^{\bar{a}_K}_t=\hat{E}(\hat{Q}^{\bar{a}_K}_{t+1}|\bar{A}_{t}=\bar{a}_t,\bar{L}_{t},C_t=0)$ using the estimated $\hat{\gamma}_0, \hat{\gamma}_1, \hat{\gamma}_2$ from step 3a
\EndFor
\State 3c. The counterfactual mean $E(Y^{\bar{a}_K})$ is estimated as $\dfrac{1}{n}\sum_{i = 1}^n\hat{Q}^{\bar{a}_K}_{1,i}$.
\State 4. Predict the trajectory category corresponding to the treatment pattern $\bar{a}_K$ using the fitted LCGA.
\EndFor
\State 5. Regress the vector of estimated counterfactual means $\hat{E}(Y^{\bar{a}_K})$ for all treatment patterns $\bar{a}_K$ on the predicted trajectory categories.
\State 6. The bootstrap is used to estimate the standard errors.
\end{algorithmic}
\end{algorithm}

Intuitively, this backward algorithm allows the probability of exposure at each time point to be adjusted only for previous covariates and exposure, thus bypassing the treatment-confounder feedback problem.

\subsubsection{Pooled longitudinal targeted maximum likelihood estimator (pooled LTMLE)}

The IPW and g-computation estimators require, respectively, the correct specification of the treatment model or the outcome model to obtain valid estimates. The pooled LTMLE method is a doubly robust estimation method that uses both outcome and treatment modelling, such that valid estimates are obtained when at least one of the model groups (i.e. either all of the treatment models or all of the outcomes models) is correctly specified \citep{petersen2014targeted, luque2018targeted}. Doubly robust methods also have the advantage that they can be implemented with machine learning techniques, instead of parametric models, for the treatment and outcome regressions. In the pooled LTMLE, the target parameter is expressed as a function of the counterfactual means, as in g-computation: 
$$E(Y^{\bar{a}_K}) = E(Q^{\bar{a}_K}_{1}).$$ 
After initial estimation, the counterfactual means at each time $t$ are updated using weights with a form close to what is used in the IPW. These weights are computed for all the $2^K$ treatment patterns simultaneously; these weights are estimates of:
$$w_t = \dfrac{1}{\prod_{j=1}^{t}P(A_{j}=a_j|\bar{A}_{j-1}=\bar{a}_{j-1},\bar{L}_{j},C_{j-1}=0)P(C_j=0\mid \bar{A}_j=\bar{a}_j, \bar{L}_j,C_{j-1}=0)}.$$
An update matrix is also computed at each time:
$$\textbf{H}_t=I(\bar{A}_{t}=\bar{a}_{t},C_t=0)\dfrac{\partial}{\partial \beta}m(\bar{a}_{t}|\beta)Var^{-1}(\bar{a}_{t}|\beta).$$
Given an estimated vector $\hat{Q}^{\bar{a}_{K}^*}_{t+1}$ from the previous iteration, and an initial estimate $\hat{Q}^{\bar{a}_{K}}_{t}=\hat{E}(\hat{Q}^{\bar{a}_{K}^*}_{t+1}\mid \bar{A}_t=\bar{a}_t,\bar{L}_t, C_t=0)$, the fluctuation vector  $\epsilon$ is estimated using $w_t, t =1 \ldots, K$, as weights and $\textbf{H}_t$ as covariates \citep{luque2018targeted, tran2019double}:
$$\text{logit}(\hat{Q}^{\bar{a}_{K}^*}_{t+1})=\text{logit}(\hat{Q}^{\bar{a}_{K}}_{t})+\bm{\epsilon}\textbf{H}_{t},$$
where $\hat{\bm \epsilon}$ minimizes the risk under a logistic loss function.
At each time $t$ and for all $2^K$ treatment patterns, the counterfactual mean is updated with the following expression:
$$\hat{Q}^{\bar{a}_{K}^*}_{t}=\text{expit}(\text{logit}(\hat{Q}^{\bar{a}_K}_t)+\hat{\bm{\epsilon}}\textbf{H}_t).$$
The efficient influence function (EIF) is used to estimate the standard errors. When the observations are independent and identically distributed, the variance of the LTMLE is given by:
$$var(\hat{\psi})\approx var\left(\dfrac{1}{n}\sum_{i=1}^{n}IF_{i,\beta}\right)=\dfrac{1}{n}var\left(IF_{i,\beta}\right).$$
The unfamiliar and interested reader may refer to \citet{luque2018targeted} for a tutorial on TMLE and to \citet{petersen2014targeted} for a detailed presentation of the pooled LTMLE algorithm. 

\section{LCGA and History-Restricted MSMs (LCGA-HRMSMs)}

In LCGA-MSM, we only considered one time interval to model the trajectories, and the outcome was measured at the end of the follow-up period of the treatment. This approach can lead to inefficient data usage as we have to exclude individuals who experienced the outcome during the time interval used to determine the trajectory groups. Additionally, we ignore what happens to the trajectories during the follow-up period of the outcome. To address these limitations and make the most of the data, we introduced a combination of history-restricted MSMs (HRMSMs) and LCGA \citep{diop2024history}. HRMSMs are a generalization of standard MSMs where the complete follow-up period is divided into multiple shorter time intervals. The outcome is then measured at the end of each time interval, allowing for a time-dependent outcome to be taken into account. Let $s$ denote the length of a time interval. When the total follow-up time is $K$, we can create $K-s+1$ subsets. We index the time intervals by $d  = 1,\ldots, K-s+1$. These multiple subsets are considered simultaneously when estimating the trajectory groups. The procedure involves applying LCGA to the pooled data, where the time variable consists of the time intervals $d$. The identifier for individuals accounts for the repeated measurements within each subset $d$. More simply, LCGA-HRMSM can be viewed as a repeated application of LCGA-MSM at different time intervals. 

We present the case of a time-to-event outcome which is currently supported in the package \pkg{trajmsm}. Thus, it seems natural to define the nonparametric HRMSM as a model for hazard ratios (HRs). However, HRs lead to interpretation problems notably due to the built-in selection bias of hazard ratios and the impossibility to map the trajectory groups into a single hazard ratio \citep[see][for more details]{hernan2010hazards, neugebauer2007causal}. Thus, to circumvent these challenges, modelling the absolute risk is a better alternative. To model the absolute risk, the nonparametric HRMSM with a time-dependent outcome for the d$th$ interval $I_d = [d,d+s-1]$ is defined as:

\begin{equation}
    E_{Y_{d+s}^{\bar{a}_d}}(Y_{d+s}^{\bar{a}_d}| Y_d = 0) = m^*(\bar{a}_d|Y_d = 0).
\end{equation}

The causal parameter of interest $\beta$ represents the vector that minimizes the statistical distance between the true causal marginal risk and the the marginal risk as defined by a working model \citep{diop2024history}. Except for the intercept, the $\beta$ parameters are not time-dependent and represent a pooled effect over the time-intervals. 
To model the absolute risks, a natural choice of an HRMSM working model is to consider a log-linear model:
$$\log\{E(Y^{\bar{a}_d} = 1|Y_{t-1 = 0})\}
=\beta_{0,d}+\beta_1z^*_1(\bar{a}_d)+\beta_2z^*_2(\bar{a}_d)+\ldots+\beta_{J-1}z^*_{J -1}(\bar{a}_d).$$ 

Two common options for estimating the parameters of this log-linear model are the log-binomial and the Poisson loss functions.

\subsection{Estimation of the parameters of LCGA-HRMSMs}

We propose estimating the vector of LCGA-HRMSM parameters $\beta_{0,d}, \beta_1, \ldots, \beta_{J-1}$ using IPW, g-computation, and pooled LTMLE. The estimation procedure follows the approach described in Section \ref{s:estim}.

\subsubsection{IPW Estimation}

The IPW estimator sequentially creates pseudo-populations where, at each time $t$ and interval $d$, the exposure $A_{d,t}$ is independent of confounders $\bar{L}_{d,t}$. This process involves estimating the weights, typically through pooled logistic regression across all time intervals. Next, a pooled regression model is fitted to the outcome over the time intervals, using a selected loss function such as Poisson. The trajectory group, estimated simultaneously using the pooled data, is included as a covariate to estimate absolute risks. Standard errors and confidence intervals are obtained using the sandwich variance estimator through generalized estimating equations (GEE), which accounts for the multiple contributions of individuals across different time intervals. The geeglm function from the R package \pkg{geepack} allows fitting a robust Poisson model.

\subsubsection{G-Computation}

G-computation can be applied in the context of LCGA-HRMSM by first estimating counterfactual expectations for each time interval and each treatment pattern using iterated conditional expectation algorithms. This step ensures that the expected outcome for each treatment pattern is modeled separately for each subset. Once the counterfactual expectations are computed, each deterministic treatment pattern is assigned to a trajectory group. This is done using the trajectory groups identified by the previously fitted LCGA model on the observed data. The assignment links the counterfactual outcomes to the trajectory groups.

Next, the counterfactual expectations are regressed onto the trajectory categories and time intervals using a pooled regression with either a log-binomial or Poisson model. This model allows for the estimation of absolute risks, accounting for both the latent group structures and the time-varying aspects of the data.

Finally, standard errors are calculated using block bootstrapping, which accounts for the multiple contributions of each individual across time intervals. This ensures that the variability in the repeated measures data is appropriately captured.

\subsubsection{Pooled LTMLE}

For a doubly robust approach, pooled LTMLE can be implemented. Updated counterfactual means are estimated at each interval $d$, and a pooled regression model is used to estimate the parameter of interest $\beta$. We propose a variance estimator correction based on the efficient influence function to account for repeated observations \citep{schnitzer2014effect}.

$$var(\hat{\psi})\approx var\left(\dfrac{1}{n}\sum_{d\in \mathcal{T}_{s}}\sum_{i=1}^{n_d}IF_{i,d,\beta}\right).$$ 
The final form of the variance is expressed as follows:
\begin{equation}
\sigma^2=(1/n^2)\left[\sum_{d\in \mathcal{T}_{s}}n_d\sigma_d^2+2\sum_{d\in \mathcal{T}_{s}}\sum_{d^{'}>d}n_{d^{'}}\rho_{d,d^{'}}\right],
\end{equation}

where $\mathcal{T}_s$ is the set indexing all time intervals, $n_d$ represents the number of individuals at time interval $d$, $\sigma_d^2=var(IF_{i,d, \beta})$ is the variance of the efficient influence function in each of the $d=1, \ldots, K-s+1$ subsets for the $i$th individual, and $\rho_{d,d^{'}}=cov(IF_{i,d},IF_{i,d^{'}, \beta})$ denotes the within-subject correlation between influence functions at different intervals $d$ and $d^{'}$. This formulation takes into account both the within-interval variance and the correlation between intervals, providing a comprehensive estimate of the overall variance.

\section{Presentation of the trajmsm package}

In this section, we present the features of the package \pkg{trajmsm} version 0.1.3. We also present how to use the main functions of the package with examples. The functions for LCGA-MSM are designed to facilitate  estimation while providing the user with enough flexibility to build and visualize the trajectory groups. In contrast, for the more advanced LCGA-HRMSM, the function manages all the steps from the construction of the trajectory groups to the estimation of the parameters. 

\subsection{Features of the trajmsm package}

The current version of the \pkg{trajmsm} package has a total of 10 functions (see Table \ref{features}). The first 6 functions of the package allow for the estimation of the parameters of LCGA-MSMs and LCGA-HRMSMs using the three following estimation methods: IPW, g-computation and pooled LTMLE. Other functions are available to build and visualize the trajectory, generate simulated data in order to test the functions, and to split the data into different time intervals.
\begin{table}[htbp]
\small
  \centering
  \caption{Features of the R \pkg{trajmsm} package}
    \begin{tabular}{lll}
    \hline
    \toprule
    \multicolumn{1}{c}{} & \multicolumn{1}{c}{ \textbf{Name}} & \multicolumn{1}{c}{\textbf{Function}} \\
    \midrule
    \hline
    1     & \code{trajmsm\_ipw} & For estimating parameters of LCGA-MSMs using IPW.\\
    2     & \code{trajmsm\_gform}       & For estimating parameters of LCGA-MSMs using g-computation. \\
    3     & \code{trajmsm\_pltmle}      &  For estimating parameters of LCGA-MSMs using pooled LTMLE.\\
    4     &     \code{trajhrmsm\_ipw}  &  For estimating parameters of LCGA-HRMSMs using IPW.  \\
    5     &  \code{trajhrmsm\_gform}     &   For estimating parameters of LCGA-HRMSMs using g-computation. \\
    6     &    \code{trajhrmsm\_pltmle}    & For estimating parameters of LCGA-HRMSMs using pooled LTMLE.  \\
    7     &   \code{build\_traj}   &  To summarize the time-varying treatment into few groups.   \\
    8     &   \code{ggtraj}   &   To visualize the trajectory groups.  \\
    9     &   \code{gendata}    &   To generate simulated datasets. \\
    10    &   \code{split\_data}    & To split the data into multiple subsets of size $s$.    \\
    \bottomrule
    \hline
    \end{tabular}
  \label{features}
\end{table}

\subsection{LCGA-MSM with the trajmsm package}

In this section, we present the different steps required for implementing an LCGA-MSM analysis using our proposed package. To illustrate the process, we first generate simulated data using the \code{gendata} function (see Table \ref{ref_samp_data}). These generated data are in wide format (one row per subject) and are comprised of information on baseline covariates $age$ and $sex$, the time-varying treatment $statins$, time-varying covariates $hypertension$ and $bmi$, and an outcome $Y$ measured at the end of the follow-up period.

\begin{lstlisting}[style=customr]
library(trajmsm)
obsdata <- gendata(n = 1000, format = "wide", start_year = 2011, 
           total_followup = 3, seed = 345)
head(obsdata)
\end{lstlisting}

\begin{table}[h!]
\centering
\caption{Sample data in a wide format for LCGA-MSMs}
    \resizebox{.95\textwidth}{!}{
\begin{tabular}{rrrrrrrrrrrrrr}
  \hline
 & id & age & sex & y & statins2011 & statins2012 & statins2013 & hyper2011 & hyper2012 & hyper2013 & bmi2011 & bmi2012 & bmi2013 \\ 
  \hline
&   1 &   0 &   0 &   0 &   1 &   1 &   1 &   1 &   1 &   1 &   0 &   1 &   0 \\ 
 &   2 &   0 &   0 &   0 &   1 &   0 &   0 &   1 &   0 &   1 &   1 &   1 &   0 \\ 
  &   3 &   0 &   1 &   0 &   0 &   0 &   1 &   1 &   1 &   1 &   0 &   1 &   1 \\ 
   &   4 &   1 &   0 &   0 &   1 &   1 &   1 &   0 &   1 &   0 &   1 &   1 &   0 \\ 
   &   5 &   0 &   0 &   0 &   0 &   0 &   1 &   1 &   0 &   0 &   1 &   0 &   1 \\ 
  &   6 &   1 &   0 &   1 &   0 &   0 &   1 &   1 &   0 &   1 &   0 &   0 &   1 \\ 
   \hline
\end{tabular}}
\label{ref_samp_data}
\end{table}

\subsubsection{Identification of the trajectory groups }\label{s:iden_traj}

The next step is to identify the trajectory groups using the function \code{build\_traj}. This function allows to model the treatment and summarize the different treatment patterns into few trajectory groups. It is a wrapper of the \code{flexmix} function from the \proglang{R} package \pkg{FlexMix} \citep{leisch2004flexmix}. However, before using the \code{build\_traj} function, the data must first be converted to long format. This can be achieved using the \proglang{R} base function \code{reshape}. Table (\ref{samp_data_lcga}) shows the data format after reshaping. It is noteworthy that the function \code{gendata} can generate data directly in long format using the \code{format} option.

\begin{lstlisting}[style=customr]
obsdata_long = reshape(data = obsdata,
                       varying = list(c(5, 8, 11), c(6, 9, 12), c(7, 10, 13)),
                       v.names = c("statins", "hyper", "bmi"),
                       idvar = "id", times = 2011:2013, direction = "long")
head(obsdata_long)
\end{lstlisting}

\begin{table}[h!]
\centering
\caption{Format of the data used to build the trajectory groups}
\begin{tabular}{rrrrrrrrrr}
  \hline
 & id & age & sex & y & time & statins & hyper & bmi & censor \\ 
  \hline
 & 1 & 0 & 0 & 1 & 2011 & 1 & 1 & 1 & 0 \\ 
  & 2 & 0 & 1 & 1 & 2011 & 0 & 0 & 1 & 0 \\ 
 & 3 & 0 & 0 & 1 & 2011 & 1 & 1 & 0 & 0 \\ 
   & 4 & 1 & 1 & 0 & 2011 & 1 & 1 & 1 & 0 \\ 
   & 5 & 0 & 0 & 1 & 2011 & 1 & 1 & 0 & 0 \\ 
& 6 & 1 & 1 & 1 & 2011 & 1 & 1 & 1 & 0 \\ 
   \hline
\end{tabular}
\label{samp_data_lcga}
\end{table}

To assign individuals to a group, the \code{build\_traj} function uses the largest posterior probability criterion (Equation \ref{eq2}) as suggested by \citet{nagin2005group}. The function \code{build\_traj} currently supports binary treatments. Four arguments are required: the data in a long format with the argument \proglang{Rdat}, the number of trajectories with the argument  \code{number\_traj}, the formula to define the polynomial form to model the trajectories and the identifier (id) of the individuals. The \code{build\_traj} function gives as output the estimated model and the posterior probabilities of each trajectory group. For example, to fit an LCGA with \code{number\_traj} $ = 3$ trajectory groups where the logit of the probability of treatment is a linear function of time:

\begin{lstlisting}[style=customr]
restraj = build_traj(obsdata = obsdata_long, number_traj = 3, 
          formula = cbind(statins,1-statins) ~ time , identifier = "id")
\end{lstlisting}

We can also extract the BIC of the estimated model from the \code{build\_traj} object: 

\begin{lstlisting}[style=customr]
BIC(restraj$traj_model)
[1] 3558.379
\end{lstlisting}

As another example, to fit a model with \code{number\_traj} $= 3$ trajectory groups of quadratic order, the following code can be used:

\begin{lstlisting}[style=customr]
restraj = build_traj(obsdata = obsdata_long, number_traj = 3, 
          formula = cbind(statins,1-statins) ~ time + I(time^2), identifier = "id")
\end{lstlisting}

As previously mentioned, multiple models with different number of groups and varying polynomial order would typically be fitted at this stage of the analysis. A model would be selected among those fitted based on the BIC (where a lower BIC is preferred) together with substantive knowledge and interpretability of the solution. Visualizing the trajectory groups (next section) can help to assess whether a given model is reasonable from a substantive perspective.

\subsubsection{Visualization of the trajectory groups}

The \code{ggtraj} function, built using the \pkg{ggplot2} package, allows to visualize the observed treatment over time for each trajectory group. We can choose how to summarize the treatment data at each time point. For example, for a binary treatment, the function summarizes the data by calculating the proportion of individuals receiving the treatment at each time point during the follow-up period. These proportions are then displayed over time and for each trajectory group using a ggplot figure. This function requires the following inputs:
\begin{lstlisting}[style=customr]
ggtraj(traj_data, treatment, time, identifier, class, FUN = mean)
\end{lstlisting}

\begin{itemize}
    \item Time-varying treatment data in long format (``traj\_data''), including the trajectory groups, which can be extracted from the output of the \texttt{build\_traj} function.
    \item Names of the treatment, time, identifier, and trajectory group variables (``class").
    \item A formula to compute a function of the observed treatment at each time point (e.g., ``FUN = mean'').
\end{itemize}

By default, \code{ggtraj} computes the mean of the observed treatment over time. Returning to the example, we first extracted the trajectory groups from the output of \texttt{build\_traj}, which was previously saved in the object \texttt{restraj}, and saved it in the object \texttt{datapost}.

\begin{lstlisting}[style=customr]
datapost = restraj$data_post
head(datapost)
\end{lstlisting}

\begin{table}[ht]
\centering
\caption{Observed treatment Probabilities and Classification}
\begin{tabular}{rrrrlr}
  \hline
 & X1 & X2 & X3 & class & id \\ 
  \hline
1 & 0.20 & 0.34 & 0.46 & 3 &   1 \\ 
  2 & 0.02 & 0.17 & 0.81 & 3 &   2 \\ 
  3 & 0.20 & 0.34 & 0.46 & 3 &   3 \\ 
  4 & 0.70 & 0.22 & 0.08 & 1 &   4 \\ 
  5 & 0.70 & 0.22 & 0.08 & 1 &   5 \\ 
  6 & 0.70 & 0.22 & 0.08 & 1 &   6 \\ 
   \hline
\end{tabular}
\label{postprob}
\end{table}

To produce a plot using the \code{ggtraj} function, we then merged \texttt{datapost} with the observed data. We then calculated the proportion (probability) of treatment for each group at each time point over the follow-up period. Next, we relabeled the classification variable in decreasing order of observed treatment probability (from high to low). This new data frame is used to plot the observed treatment probabilities for each trajectory group over time.

\begin{lstlisting}[style=customr]
traj_data_long <- merge(obsdata_long, datapost, by = "id")
     AggFormula <- as.formula(paste("statins", "~", "time", "+", "class"))
     Aggtraj_data <- aggregate(AggFormula, data = traj_data_long, FUN = mean)
     Aggtraj_data
     
 #Aggtraj_data with labels
traj_data_long[ , "traj_group"] <- 
factor(ifelse(traj_data_long[ , "class"] == "2" ,"Group1" , 
ifelse (traj_data_long[ , "class"]== "1" , "Group2" ,"Group3")))
\end{lstlisting}

\begin{table}[ht]
\centering
\caption{Mean of observed treatment probabilities per trajectory groups and time points}
\begin{tabular}{rrlr}
  \hline
 & time & traj\_group & statins \\ 
  \hline
1 & 2011 & Group1 & 1.00 \\ 
  2 & 2012 & Group1 & 1.00 \\ 
  3 & 2013 & Group1 & 0.00 \\ 
  4 & 2011 & Group2 & 0.78 \\ 
  5 & 2012 & Group2 & 1.00 \\ 
  6 & 2013 & Group2 & 1.00 \\ 
  7 & 2011 & Group3 & 0.48 \\ 
  8 & 2012 & Group3 & 0.16 \\ 
  9 & 2013 & Group3 & 0.48 \\ 
   \hline
\end{tabular}
\label{adhmean}
\end{table}
\newpage
In the example below, we plot a graph of the proportion of the treatment variable at each time point for each trajectory group. To facilitate interpretation, we first rename the trajectory groups based on the observed treatment probabilities, from the highest (group 1) to the lowest (group 3), before using the \code{ggtraj} function.

\begin{lstlisting}[style=customr]
obsdata_long = gendata(n = 1000, format = "long", total_followup = 12, seed = 945)
restraj = build_traj(obsdata = obsdata_long, number_traj = 3, 
formula = as.formula(cbind(statins, 1 - statins) ~ time), identifier = "id")
datapost = restraj$data_post
head(datapost)
traj_data_long <- merge(obsdata_long, datapost, by = "id")
    AggFormula <- as.formula(paste("statins", "~", "time", "+", "class"))
    Aggtraj_data <- aggregate(AggFormula, data = traj_data_long, FUN = mean)
    Aggtraj_data
#Aggtraj_data with labels
traj_data_long[ , "traj_group"] <- factor(ifelse(traj_data_long[ , "class"] == "2" ,
"Group3",ifelse (traj_data_long[ , "class"]== "3" , "Group1" ,"Group2")))

AggFormula <- as.formula(paste("statins", "~", "time", "+", "traj_group"))
Aggtraj_data <- aggregate(AggFormula, data = traj_data_long, FUN = mean)
ggtraj(traj_data =  Aggtraj_data,
treatment = "statins",time= "time",identifier="id",class = "traj_group", FUN = mean)
\end{lstlisting}

\begin{figure}[h!]
    \centering
    \includegraphics[scale = 0.5]{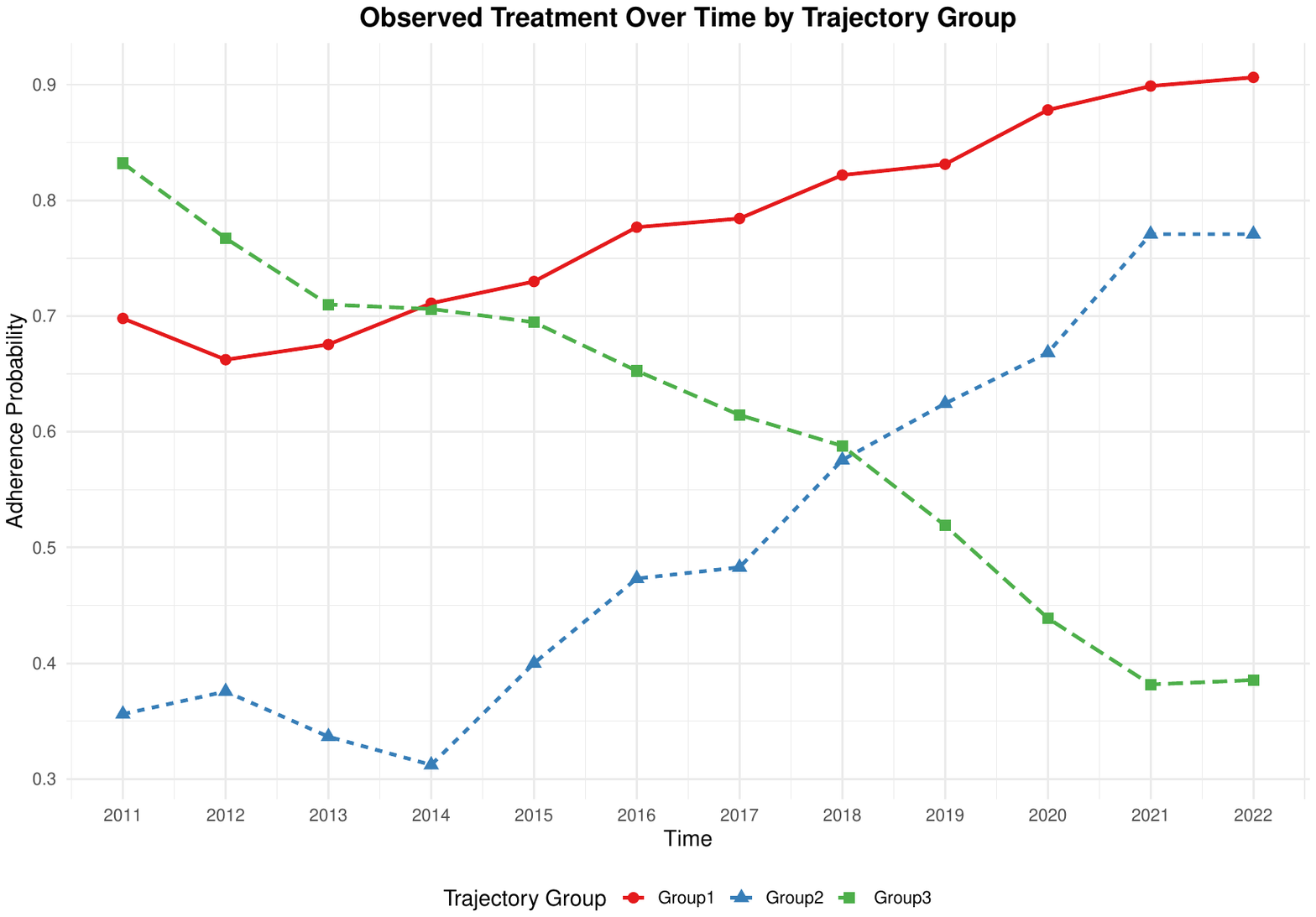}
    \caption{Trajectory groups for 12 follow-up periods }
    \label{fig_tp1b}
\end{figure}

Figure \ref{fig_tp1b} depicts the result of the previous code for 12 follow-up periods.

\subsubsection{Estimating the parameters of an LCGA-MSM using IPW}

The function \code{trajmsm\_ipw} estimates the parameters of an LCGA-MSM using IPW. It provides the estimates, standard errors, confidence intervals, and p-values for the weighted regression of the outcome of interest on the trajectory groups. The function \code{trajmsm\_ipw} accounts for potential time-varying confounders as well as treatment-confounder feedback when estimating the treatment effect.

This function accepts several arguments:
\begin{itemize}
    \item \code{formula1}: Specifies a formula for modeling the outcome when it is continuous or binary.
    \item \code{formula2}: Specifies a model for time-to-event outcomes.
    \item \code{numerator}: Chooses between unstabilized or stabilized weights.
    \item \code{identifier}: Specifies the identifier for individuals.
    \item \code{treatment}: Specifies the name of the treatment variable.
    \item \code{covariates}: Specifies the names of the time-varying covariates.
    \item \code{baseline}: Specifies the names of baseline variables.
    \item \code{total\_followup}: Specifies the length of the follow-up period.
    \item \code{number\_traj}: Specifies the number of trajectory groups.
    \item \code{include\_censor}: Indicates whether there is censoring or not.
    \item \code{censor}: Specifies the name of the censoring variable, if applicable.
    \item \code{obsdata}: Specifies the observed data in wide format.
\end{itemize}

\begin{lstlisting}[style=customr]
obsdata_long = gendata(n = 5000, format = "long", total_followup = 6, seed = 845)
years <- 2011:2016
baseline_var <- c("age","sex")
variables <- c("hyper", "bmi")
covariates <- lapply(years, function(year) {paste0(variables, year)})
treatment_var <- paste0("statins", 2011:2016)
formula_treatment = as.formula(cbind(statins, 1 - statins) ~ time)
restraj = build_traj(obsdata = obsdata_long, number_traj = 3,
    formula = formula_treatment, identifier = "id")
datapost = restraj$data_post
trajmsm_long <- merge(obsdata_long, datapost, by = "id")
AggFormula <- as.formula(paste("statins", "~", "time", "+", "class"))
AggTrajData <- aggregate(AggFormula, data = trajmsm_long, FUN = mean)
trajmsm_long$ipw_group <- relevel(trajmsm_long$class, ref = "1")
obsdata = reshape(data = trajmsm_long, direction = "wide", idvar = "id",
    v.names = c("statins","bmi","hyper"), timevar = "time", sep ="")

resmsm_ipw = trajmsm_ipw(formula1 = as.formula("y ~ ipw_group"),
    identifier = "id", baseline = baseline_var, covariates = covariates,
    treatment = treatment_var, number_traj=3, total_followup = 6, family = "binomial",
    obsdata = obsdata,numerator = "stabilized", include_censor = FALSE)
resmsm_ipw   
\end{lstlisting}

Prior to estimation, the reference group should be defined, for example, using the \proglang{R} base function \code{relevel}.  Given that the outcome is binary, a binomial distribution with a logit link function was employed. Table (\ref{est_lcga_msm_ipw}) summarizes the parameter estimates ($\beta$), standard errors, p-values, and 95\% confidence intervals for the LCGA-MSM model using IPW. 

Recall that these results are derived from simulated data and do not reflect real-world findings. In this example, the outcome variable $y$ represents a cardiovascular event, where a value of 1 indicates the occurrence of a cardiovascular event and 0 indicates no event. The highest overall adherence group (Group 1) serves as the reference, showing relatively low odds of experiencing a cardiovascular event. More precisely, the odds for this group are calculated as $ \exp(-1.66) \approx 0.19$, meaning the odds of experiencing a cardiovascular event are 0.19. This suggests that individuals in this group have a relatively low chance of experiencing such an event. The results further suggest that individuals with an adherence that is very low at first but that increases over time have (Group 2) an even lower odds of cardiovascular events, with an odds ratio (OR) of $\exp(-1.46) \approx 0.23$. This means that individuals in this ``increasing adherence'' group have about 77\% lower odds of experiencing a cardiovascular event compared to those in the highest overall adherence group. Individuals with an adherence that is very high at first but that decreases over time (Group 3) also have moderately lower odds of cardiovascular events than those in the highest overall adherence (Group 1), with an OR = 0.74. 


\begin{table}[h!]
\centering
\caption{Estimation of the parameters of an LCGA-MSM using IPW with stabilized weights}
\begin{tabular}{rrrrrr}
  \hline
 & Estimate & Std.Error & Pvalue & Lower CI & Upper CI \\ 
  \hline
(Intercept) & -1.66 & 0.09 & 0.00 & -1.84 & -1.49 \\ 
  ipw\_group2 & -1.46 & 0.13 & 0.00 & -1.72 & -1.21 \\ 
  ipw\_group3 & -0.30 & 0.13 & 0.02 & -0.56 & -0.05 \\ 
   \hline
\end{tabular}
\label{est_lcga_msm_ipw}
\end{table}

\subsubsection{Estimating the parameters of an LCGA-MSM with g-computation}

As discussed in Section \ref{s:estim}, estimation with g-computation in the context of LCGA-MSM involves four key steps: (1) estimating the counterfactual means, (2) assigning treatment patterns to the trajectory groups, (3) regressing the counterfactual means onto the trajectory groups, and (4) performing inference using a bootstrapping approach to quantify uncertainty. The function \code{trajmsm\_gform} facilitates this process by taking the following inputs:

\begin{itemize}
    \item \code{formula}: The formula to model the counterfactual means.
    \item \code{trajmodel}: The fitted LCGA model.
    \item \code{identifier}: The name of the individual identifier variable.
    \item \code{baseline}: The names of the baseline covariates.
    \item \code{covariates}: The names of the time-varying covariates.
    \item \code{treatment}: The name of the treatment variable.
    \item \code{outcome}: The name of the outcome variable.
    \item \code{ref}: The reference category.
    \item \code{total\_followup}: The length of the follow-up period.
    \item \code{time}: The name of the time variable.
    \item \code{time\_values}: The values of the time variable.
    \item \code{rep}: The number of repetitions for the bootstrap (default is 50, but a higher number, such as 200, is recommended).
\end{itemize}

The results are presented in Table \ref{est_lcga_msm_gform} and include LCGA-MSM parameter estimates, standard errors, p-values, and 95\% confidence intervals.

\begin{lstlisting}[style=customr]
v.names = c("statins","bmi","hyper"), timevar = "time", sep ="")
formula = paste0("y ~", paste0(treatment_var,collapse = "+"), "+",
    paste0(unlist(covariates), collapse = "+"),"+",
    paste0(baseline_var, collapse = "+"))
resmsm_gform <- trajmsm_gform(formula = formula, identifier = "id",rep = 50,
baseline = baseline_var, covariates = covariates, var_cov = var_cov,
    treatment = treatment_var, outcome = "y", total_followup = 6,time = "time",
    time_values = years, trajmodel = restraj$traj_model,ref = "1", obsdata = obsdata )
resmsm_gform
\end{lstlisting}

\begin{table}[ht]
\centering
\caption{Estimation of the parameters of an LCGA-MSM using g-computation}
\begin{tabular}{rrrrrr}
  \hline
 & Estimate & Std.Error & Pvalue & Lower CI & Upper CI \\ 
  \hline
(Intercept) & -1.54 & 0.07 & 0.00 & -1.68 & -1.40 \\ 
  gform\_group2 & -1.32 & 0.10 & 0.00 & -1.51 & -1.13 \\ 
  gform\_group3 & -0.26 & 0.08 & 0.00 & -0.41 & -0.10 \\ 
   \hline
\end{tabular}
\label{est_lcga_msm_gform}
\end{table}
\subsubsection{Estimating the parameters of an LCGA-MSM using pooled LTMLE}

Estimation of the parameters of an LCGA-MSM with pooled LTMLE is achieved using the function \code{trajmsm\_pltmle}. This function takes the same arguments as the function \code{trajmsm\_gform}, except that we do not need to specify the number of repetitions since standard errors are not computed using bootstrap but rather the influence functions. Results are presented in Table (\ref{est_lcga_msm_pltmle}).

\begin{lstlisting}[style=customr]
trajmsm_pltmle(formula = formula, identifier = identifier, baseline = baseline,
    covariates = covariates, treatment = treatment, outcome = "y", time = "time",
    time_values = time_values, number_traj = 3, total_followup = 3,
    trajmodel = restraj$traj_model, ref = "3", obsdata = trajmsm_wide)
\end{lstlisting}

\begin{table}[h!]
\centering
\caption{Estimation of the parameters of an LCGA-MSM using pooled LTMLE}
\begin{tabular}{rrrrrr}
  \hline
 & Estimate & Std.Error & Pvalue & Lower CI & Upper CI \\ 
  \hline
(Intercept) & -1.60 & 0.04 & 0.00 & -1.68 & -1.52 \\ 
  ptmle\_group2 & -1.28 & 0.15 & 0.00 & -1.58 & -0.98 \\ 
  ptmle\_group3 & -0.27 & 0.08 & 0.00 & -0.43 & -0.11 \\ 
   \hline
\end{tabular}
\label{est_lcga_msm_pltmle}
\end{table}

\subsection{LCGA in the context of HRMSM}

To begin, we generate a dataset that includes a sample of individuals over a specified follow-up period with baseline characteristics (e.g., age, sex), treatment variables (e.g., statins), and covariates (e.g., hypertension, BMI). Afterward, we segment the data into multiple time intervals using the \code{split\_data} function.The function \code{split\_data} converts the data into the appropriate format for LCGA-HRMSM analysis. The inputs for this function are:

\begin{itemize}
    \item The observed data in long format.
    \item The total length of follow-up.
    \item The size of each time interval specified with \code{ntimes\_interval}.
    \item The name of the time variable and its values.
\end{itemize}

The output is a list containing data for each time interval (see Table \ref{samp_data_lcga_hrmsm}). 
In this table, a new variable \code{identifier2} is created by appending the follow-up period to the original \code{identifier}. The \code{Interv} column indicates the time interval number.

\begin{lstlisting}[style=customr]
obsdata = gendata(n = 5000, format = "long", total_followup = 8, seed = 945)
years <- 2011:2018
res = split_data(obsdata = obsdata, total_followup = 8, ntimes_interval = 6, time = "time", 
    time_values = years, identifier = "id")
head(res[[1]], n = 10)
\end{lstlisting}

\begin{table}[ht]
\centering
\caption{Sample data in a long-format for LCGA-HRMSMs}
\begin{tabular}{rrrrrrrrrrr}
  \hline
id & time & statins & hyper & bmi & age & sex & y & Interv & time2 & identifier2 \\ 
  \hline
  1 & 2011 &   1 &   0 &   0 &   1 &   0 &   0 &   1 &   1 &  1\_1 \\ 
    2 & 2011 &   0 &   1 &   1 &   1 &   0 &   0 &   1 &   1 &  2\_1 \\ 
    3 & 2011 &   1 &   0 &   1 &   1 &   0 &   0 &   1 &   1 &  3\_1 \\ 
    4 & 2011 &   1 &   0 &   0 &   1 &   1 &   0 &   1 &   1 &  4\_1 \\ 
    5 & 2011 &   1 &   0 &   0 &   0 &   0 &   0 &   1 &   1 &  5\_1 \\ 
    6 & 2011 &   1 &   0 &   0 &   1 &   0 &   0 &   1 &   1 &  6\_1 \\ 
    7 & 2011 &   0 &   0 &   1 &   0 &   0 &   0 &   1 &   1 &  7\_1 \\ 
    8 & 2011 &   1 &   1 &   1 &   0 &   1 &   0 &   1 &   1 &  8\_1 \\ 
    9 & 2011 &   0 &   0 &   0 &   1 &   1 &   0 &   1 &   1 &  9\_1 \\ 
   10 & 2011 &   1 &   0 &   1 &   1 &   0 &   0 &   1 &   1 & 10\_1 \\ 
   \hline
\end{tabular}
\label{samp_data_lcga_hrmsm}
\end{table}

\begin{figure}[h!]
    \centering
    \includegraphics[scale = 0.5]{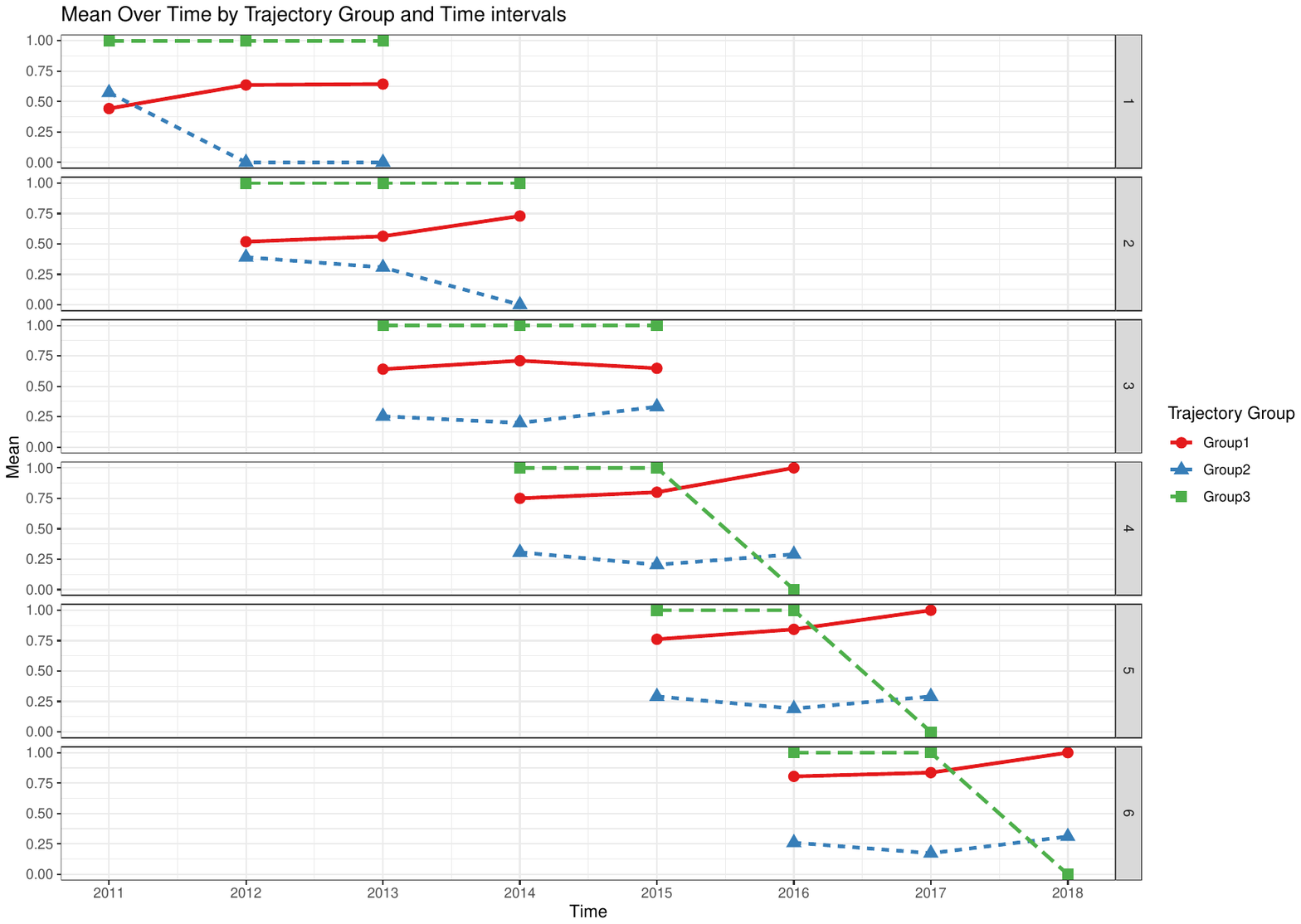}
    \caption{Mean treatment levels over time by trajectory group}
    \label{fig_tp2b}
\end{figure}

Once the data is split, we apply a trajectory model on the pooled data to cluster individuals based on observed treatment patterns, which helps to understand how responses to treatment evolve over time. Following this, we merge the trajectory groups with the augmented dataset. After merging, we assign labels to the trajectory groups for easier identification in subsequent analyses. We compute the proportion of treatment by time interval and trajectory group, which provides insight into variations in treatment outcomes across different classes over time. Finally, we visualize the results by creating plots that illustrate mean treatment levels over time for each trajectory group, using facets to display results by interval, highlighting differences in treatment responses among the groups (see Figure \ref{fig_tp2b}).

\begin{lstlisting}[style=customr]
# Step 1: Generate a synthetic dataset with 1000 individuals over 8 years
obsdata_long = gendata(n = 1000, format = "long", total_followup = 8, 
                       timedep_outcome = TRUE, seed = 845)
baseline_var <- c("age", "sex") 
years <- 2011:2018 
variables <- c("hyper", "bmi") 
covariates <- lapply(years, function(year) { paste0(variables, year) }) 
treatment_var <- paste0("statins", 2011:2018) 
var_cov <- c("statins", "hyper", "bmi", "y") 

# Step 2: Transform the dataset into different time intervals
dat_sub = data.frame(do.call(rbind, split_data(obsdata = obsdata_long, 
                                               total_followup = 8, 
                                               ntimes_interval = 3, 
                                               time = "time", 
                                               time_values = years, 
                                               identifier = "id")))

# Step 3: Build trajectory groups based on the treatment pattern
formula_treatment = as.formula(cbind(statins, 1 - statins) ~ time)
restraj = build_traj(obsdata = dat_sub, number_traj = 3, 
                     formula = formula_treatment, 
                     identifier = "identifier2")

# Step 4:Merge trajectory groups with the augmented dataset
datapost = restraj$data_post 
head(datapost) 
traj_data_long <- merge(dat_sub, datapost, by = "identifier2") 

# Step 5: Aggregate the treatment data by time and trajectory groups
AggFormula <- as.formula(paste("statins", "~", "time", "+", "class", "+ Interv")) 
Aggtraj_data <- aggregate(AggFormula, data = traj_data_long, FUN = mean) 
Aggtraj_data 

# Step 6: Assign labels to trajectory groups for easier identification
traj_data_long[, "traj_group"] <- 
factor(ifelse(traj_data_long[, "class"] == "2", "Group1", 
ifelse(traj_data_long[, "class"] == "1", "Group2", "Group3"))) 

# Step 7: Aggregate the treatment data again to compute means by trajectory group
# Then, create a visualization of mean treatment levels over time by trajectory group
AggFormula <- as.formula(paste("statins", "~", "time", "+", "traj_group", "+", "Interv")) 
Aggtraj_data <- aggregate(AggFormula, data = traj_data_long, FUN = mean) 
ggtraj(traj_data = Aggtraj_data, 
       treatment = "statins", time = "time", 
       identifier = "identifier2", class = "traj_group", FUN = mean) + 
  facet_grid(Interv ~ .) + 
  labs(title = "Mean Over Time by Trajectory Group and Time Intervals") +  
  theme_bw()
\end{lstlisting}

\subsection{Estimation of the parameters of an LCGA-HRMSM}

Three functions are provided to estimate the parameters for an LCGA-HRMSM: \code{trajhrmsm\_ipw}, \code{trajhrmsm\_gform}, and \code{trajhrmsm\_pltmle}. The main arguments for these functions include:

\begin{itemize}
    \item \code{degree\_traj}: The polynomial degree (linear, quadratic, or cubic) for constructing the trajectory groups.
    \item Variable names: \code{treatment}, \code{baseline}, \code{covariates}, and \code{outcome}.
    \item \code{family}: The working model (Poisson or binomial with \code{link = "log"}).
    \item \code{obsdata}: The observed data in long format.
    \item \code{numerator}: Choice between stabilized and unstabilized weights (IPW only).
    \item \code{ntimes\_interval}: The size of the time intervals.
    \item \code{total\_followup}: The total length of the follow-up period.
    \item \code{number\_traj}: The number of trajectory groups.
    \item \code{class\_var}: Name of the trajectory group variable.
\end{itemize}

The reference group is chosen based on the lowest observed treatment proportion and is directly handled by the algorithm. The output includes the results of the trajectory group estimation, the proportion of the treatment across all groups and time intervals, and the estimated parameters of the LCGA-HRMSM along with their standard errors and 95\% confidence intervals. In the case of g-computation, the vector of coefficients resulting from the block-bootstrapping is also provided. Currently, the package supports a binary time-dependent outcome and a binary time-varying exposure. For all the functions, the names of the variables should be specified. The \code{var\_cov} argument indicates the time-varying variables.

\subsubsection{Estimating the parameters of an LCGA-HRMSM using IPW}
 
The function \code{trajhrmsm\_ipw} estimates the parameters of LCGA-HRMSMs using IPW. Below is an example where the \code{total\_followup} $= 8$ time points are split in three subsets of size \code{ntimes\_interval} $= 6$ each (1-6, 2-7, 3-8) and  \code{number\_traj} $= 3$ groups are formed. 

\begin{lstlisting}[style=customr]
obsdata_long = gendata(n = 1000, format = "long", total_followup = 8, 
    timedep_outcome = TRUE,  seed = 845)
baseline_var <- c("age", "sex")
years <- 2011:2018
variables <- c("hyper", "bmi")
covariates <- lapply(years, function(year) { paste0(variables, year) } )
treatment_var <- paste0("statins", 2011:2018)
var_cov <- c("statins", "hyper", "bmi", "y)
reshrmsm_ipw <- trajhrmsm_ipw(degree_traj = "linear", numerator = "stabilized", 
    identifier = "id", baseline = baseline_var, covariates = covariates, 
    treatment = treatment_var, outcome = "y", var_cov= var_cov, include_censor = FALSE,
    ntimes_interval = 6, total_followup = 8, time = "time", time_values = 2011:2018,
    family = "poisson", number_traj = 3, obsdata = obsdata_long)
reshrmsm_ipw$res_trajhrmsm_ipw
\end{lstlisting}

In Table (\ref{avg_treat}), the observed mean of the treatment per trajectory group across all periods is shown ; such a result can be displayed using \code{res\_trajhrmsm\_ipw[[3]]}.

\begin{table}[ht]
\centering
\caption{Observed mean of the treatment per trajectory group}
\begin{tabular}{rlr}
  \hline
 & ipw\_group & statins \\ 
  \hline
& 1 & 0.25 \\ 
 & 2 & 0.63 \\ 
 & 3 & 0.76 \\ 
   \hline
\end{tabular}
\label{avg_treat}
\end{table}

In Table (\ref{est_lcga_hrmsm_ipw}), results of the estimation are displayed using the line of code: \code{reshrmsm\_ipw[[2]]}.

\begin{table}[ht]
\centering
\caption{Estimation of the parameters of an LCGA-HRMSM using IPW with stabilized weights}
\begin{tabular}{rrrrrr}
  \hline
 & Estimate & Std.Error & Pvalue & Lower CI & Upper CI \\ 
  \hline
(Intercept) & -1.15 & 0.08 & 0.00 & -1.30 & -1.00 \\ 
  factor(ipw\_group)2 & -0.18 & 0.08 & 0.02 & -0.34 & -0.03 \\ 
  factor(ipw\_group)3 & -0.30 & 0.07 & 0.00 & -0.45 & -0.16 \\ 
   \hline
\end{tabular}
\label{est_lcga_hrmsm_ipw}
\end{table}

\section{Conclusion}

The R package \pkg{{trajmsm}} facilitates the application of LCGA-MSM and LCGA-HRMSM analyses. Although our presentation focused on medical applications, this package is useful in various other fields. Indeed, trajectory analysis methods like LCGA are increasingly used in several domains, including education, criminology, and psychology. These models are relevant when the goal is to summarize numerous potential treatment trajectories into a few groups, and they can also be used to define the effects of treatment trajectory on an outcome using the LCGA-MSM methodology we have proposed.

For functions implementing LCGA-MSM analysis, the trajectory must be constructed before estimating the parameters. This approach gives users more flexibility to check the groups before estimation. For LCGA-HRMSMs, all steps are handled by the algorithms due to the procedure's complexity. However, we suggest an initial exploration of the trajectory groups before estimating the parameters of interest.

In the current version of the \pkg{{trajmsm}} package, continuous, binary, and survival outcomes are supported when estimating the parameters of LCGA-MSMs. Binary time-dependent outcomes are currently supported when estimating the parameters of LCGA-HRMSMs. In future versions of the package, we plan to include other types of outcomes for LCGA-HRMSM, such as continuous outcomes. For example, one could model the mean cholesterol level at month \(d+s\) as a function of statin use history from month \(d\) to \(d+s-1\) among those who were still alive at time \(d\). We also plan to include the bootstrap approach to estimate the standard errors as an option for all estimation methods.
\section{Appendix}

\subsubsection{Estimating parameters of LCGA-HRMSM using g-computation}

To estimate the parameters of LCGA-HRMSMs using g-computation, the function \code{trajhrmsm\_gform} is used. 
Results of the estimation are presented in Table (\ref{est_lcga_hrmsm_gform}). 

\begin{lstlisting}[style=customr]
reshrmsm_gform = trajhrmsm_gform(degree_traj = "linear", rep=50,
    treatment = treatment_var, covariates = covariates, baseline = baseline_var,
    outcome = paste0("y", 2016:2018), var_cov = var_cov, ntimes_interval = 6, 
    total_followup = 8, time = "time", time_values = years, identifier = "id",
    number_traj = 3, family = "poisson", obsdata = obsdata_long)
reshrmsm_gform$results_hrmsm_gform
\end{lstlisting}

\begin{table}[ht]
\centering
\caption{Estimation of the parameters of an LCGA-HRMSM using g-computation}
\begin{tabular}{rrrrrr}
  \hline
 & Estimate & Std.Error & Pvalue & Lower CI & Upper CI \\ 
  \hline
(Intercept) & -0.76 & 0.11 & 0.00 & -0.98 & -0.53 \\ 
  factor(gform\_group)2 & -0.13 & 0.10 & 0.19 & -0.32 & 0.06 \\ 
  factor(gform\_group)3 & -0.21 & 0.10 & 0.03 & -0.40 & -0.02 \\ 
   \hline
\end{tabular}
\label{est_lcga_hrmsm_gform}
\end{table}

\subsubsection{Estimating parameters of LCGA-HRMSM using pooled LTMLE}

Results of the estimation are presented in Table (\ref{est_lcga_hrmsm_pltmle}).
Results when using the function \code{trajhrmsm\_pltmle}:\\

\begin{lstlisting}[style=customr]
respltmle = trajhrmsm_pltmle(degree_traj = "linear", treatment = treatment_var, 
    covariates = covariates, baseline = baseline_var, outcome = paste0("y", 2016:2018), 
    var_cov = var_cov, ntimes_interval = 6, total_followup = 8, time = "time", 
    time_values = years, identifier = "id", number_traj = 3, family = "poisson", 
    obsdata = obsdata_long)
respltmle$results_hrmsm_pltmle
\end{lstlisting}

\begin{table}[ht]
\centering
\caption{Estimation of the parameters of an LCGA-HRMSM using pooled-ltmle}
\begin{tabular}{rrrrrr}
  \hline
 & Estimate & Std.Error & Pvalue & Lower CI & Upper CI \\ 
  \hline
(Intercept) & -0.70 & 0.04 & 0.00 & -0.78 & -0.61 \\ 
  factor(pltmle\_group)2 & -0.12 & 0.10 & 0.24 & -0.33 & 0.08 \\ 
  factor(pltmle\_group)3 & -0.23 & 0.08 & 0.01 & -0.39 & -0.06 \\ 
   \hline
\end{tabular}
\label{est_lcga_hrmsm_pltmle}
\end{table}

\bibliography{ref_tp}

\begin{thebibliography}{25}
\providecommand{\natexlab}[1]{#1}
\providecommand{\url}[1]{\texttt{#1}}
\expandafter\ifx\csname urlstyle\endcsname\relax
  \providecommand{\doi}[1]{doi: #1}\else
  \providecommand{\doi}{doi: \begingroup \urlstyle{rm}\Url}\fi

\bibitem[Austin(2011)]{austin2011introduction}
Peter~C Austin.
\newblock An introduction to propensity score methods for reducing the effects
  of confounding in observational studies.
\newblock \emph{Multivariate behavioral research}, 46\penalty0 (3):\penalty0
  399--424, 2011.

\bibitem[Bang and Robins(2005)]{bang2005doubly}
Heejung Bang and James~M Robins.
\newblock Doubly robust estimation in missing data and causal inference models.
\newblock \emph{Biometrics}, 61\penalty0 (4):\penalty0 962--973, 2005.

\bibitem[Cole and Hern{\'a}n(2008)]{cole2008constructing}
Stephen~R Cole and Miguel~A Hern{\'a}n.
\newblock Constructing inverse probability weights for marginal structural
  models.
\newblock \emph{American journal of epidemiology}, 168\penalty0 (6):\penalty0
  656--664, 2008.

\bibitem[Diop et~al.(2023)Diop, Sirois, Guertin, Schnitzer, Candas, Cossette,
  Poirier, Brophy, M{\'e}sidor, Blais, et~al.]{diop2023marginal}
Awa Diop, Caroline Sirois, Jason~Robert Guertin, Mireille~E Schnitzer, Bernard
  Candas, Benoit Cossette, Paul Poirier, James Brophy, Miceline M{\'e}sidor,
  Claudia Blais, et~al.
\newblock Marginal structural models with latent class growth analysis of
  treatment trajectories: Statins for primary prevention among older adults.
\newblock \emph{Statistical Methods in Medical Research}, 32\penalty0
  (11):\penalty0 2207--2225, 2023.

\bibitem[Diop et~al.(2024)Diop, Sirois, Guertin, Schnitzer, Brophy, Blais, and
  Talbot]{diop2024history}
Awa Diop, Caroline Sirois, Jason~R Guertin, Mireille~E Schnitzer, James~M
  Brophy, Claudia Blais, and Denis Talbot.
\newblock History-restricted marginal structural model and latent class growth
  analysis of treatment trajectories for a time-dependent outcome.
\newblock \emph{The International Journal of Biostatistics}, \penalty0 (0),
  2024.

\bibitem[Eisenlohr-Moul et~al.(2020)Eisenlohr-Moul, Kaiser, Weise,
  Schmalenberger, Kiesner, Ditzen, and Kleinst{\"a}uber]{eisenlohr2020there}
Tory~A Eisenlohr-Moul, Gudrun Kaiser, Cornelia Weise, Katja~M Schmalenberger,
  Jeff Kiesner, Beate Ditzen, and Maria Kleinst{\"a}uber.
\newblock Are there temporal subtypes of premenstrual dysphoric disorder?:
  using group-based trajectory modeling to identify individual differences in
  symptom change.
\newblock \emph{Psychological medicine}, 50\penalty0 (6):\penalty0 964--972,
  2020.

\bibitem[Franklin et~al.(2013)Franklin, Shrank, Pakes, Sanf{\'e}lix-Gimeno,
  Matlin, Brennan, and Choudhry]{franklin2013group}
Jessica~M Franklin, William~H Shrank, Juliana Pakes, Gabriel
  Sanf{\'e}lix-Gimeno, Olga~S Matlin, Troyen~A Brennan, and Niteesh~K Choudhry.
\newblock Group-based trajectory models: a new approach to classifying and
  predicting long-term medication adherence.
\newblock \emph{Medical care}, 51\penalty0 (9):\penalty0 789--796, 2013.

\bibitem[Gr{\"u}n and Leisch(2008)]{grun2008identifiability}
Bettina Gr{\"u}n and Friedrich Leisch.
\newblock Identifiability of finite mixtures of multinomial logit models with
  varying and fixed effects.
\newblock \emph{Journal of classification}, 25\penalty0 (2):\penalty0 225--247,
  2008.

\bibitem[Hern{\'a}n(2010)]{hernan2010hazards}
Miguel~A Hern{\'a}n.
\newblock The hazards of hazard ratios.
\newblock \emph{Epidemiology (Cambridge, Mass.)}, 21\penalty0 (1):\penalty0 13,
  2010.

\bibitem[Hernan and Robins(2020)]{hernan2010causal}
Miguel~A Hernan and James~M Robins.
\newblock \emph{Causal Inference: What If}.
\newblock Chapman and Hall/CRC Boca Raton, FL, 2020.

\bibitem[Hern{\'a}n et~al.(2000)Hern{\'a}n, Brumback, and
  Robins]{3hernan2000marginal}
Miguel~{\'A}ngel Hern{\'a}n, Babette Brumback, and James~M Robins.
\newblock Marginal structural models to estimate the causal effect of
  zidovudine on the survival of hiv-positive men.
\newblock \emph{Epidemiology}, pages 561--570, 2000.

\bibitem[Leisch(2004)]{leisch2004flexmix}
Friedrich Leisch.
\newblock Flexmix: A general framework for finite mixture models and latent
  glass regression in r.
\newblock 2004.

\bibitem[Lok(2021)]{lok2021estimating}
Judith~J Lok.
\newblock How estimating nuisance parameters can reduce the variance (with
  consistent variance estimation).
\newblock \emph{Statistics in Medicine}, 2021.

\bibitem[Luque-Fernandez et~al.(2018)Luque-Fernandez, Schomaker, Rachet, and
  Schnitzer]{luque2018targeted}
Miguel~Angel Luque-Fernandez, Michael Schomaker, Bernard Rachet, and Mireille~E
  Schnitzer.
\newblock Targeted maximum likelihood estimation for a binary treatment: A
  tutorial.
\newblock \emph{Statistics in medicine}, 37\penalty0 (16):\penalty0 2530--2546,
  2018.

\bibitem[Nagin(2005)]{nagin2005group}
Daniel~S Nagin.
\newblock \emph{Group-based modeling of development}.
\newblock Harvard University Press, 2005.

\bibitem[Nagin and Odgers(2010)]{nagin2010group}
Daniel~S Nagin and Candice~L Odgers.
\newblock Group-based trajectory modeling (nearly) two decades later.
\newblock \emph{Journal of quantitative criminology}, 26\penalty0 (4):\penalty0
  445--453, 2010.

\bibitem[Neugebauer and van~der Laan(2007)]{neugebauer2007nonparametric}
Romain Neugebauer and Mark van~der Laan.
\newblock Nonparametric causal effects based on marginal structural models.
\newblock \emph{Journal of Statistical Planning and Inference}, 137\penalty0
  (2):\penalty0 419--434, 2007.

\bibitem[Neugebauer et~al.(2007)Neugebauer, van~der Laan, Joffe, and
  Tager]{neugebauer2007causal}
Romain Neugebauer, Mark~J van~der Laan, Marshall~M Joffe, and Ira~B Tager.
\newblock Causal inference in longitudinal studies with history-restricted
  marginal structural models.
\newblock \emph{Electronic journal of statistics}, 1:\penalty0 119, 2007.

\bibitem[Petersen et~al.(2014)Petersen, Schwab, Gruber, Blaser, Schomaker, and
  {v}an~{d}er Laan]{petersen2014targeted}
Maya Petersen, Joshua Schwab, Susan Gruber, Nello Blaser, Michael Schomaker,
  and Mark {v}an~{d}er Laan.
\newblock Targeted maximum likelihood estimation for dynamic and static
  longitudinal marginal structural working models.
\newblock \emph{Journal of causal inference}, 2\penalty0 (2):\penalty0
  147--185, 2014.

\bibitem[Schnitzer et~al.(2014)Schnitzer, van~der Laan, Moodie, and
  Platt]{schnitzer2014effect}
Mireille~E Schnitzer, Mark~J van~der Laan, Erica~EM Moodie, and Robert~W Platt.
\newblock Effect of breastfeeding on gastrointestinal infection in infants: a
  targeted maximum likelihood approach for clustered longitudinal data.
\newblock \emph{The annals of applied statistics}, 8\penalty0 (2):\penalty0
  703, 2014.

\bibitem[Suerken et~al.(2016)Suerken, Reboussin, Egan, Sutfin, Wagoner,
  Spangler, and Wolfson]{suerken2016marijuana}
Cynthia~K Suerken, Beth~A Reboussin, Kathleen~L Egan, Erin~L Sutfin, Kimberly~G
  Wagoner, John Spangler, and Mark Wolfson.
\newblock Marijuana use trajectories and academic outcomes among college
  students.
\newblock \emph{Drug and alcohol dependence}, 162:\penalty0 137--145, 2016.

\bibitem[Tran et~al.(2019)Tran, Yiannoutsos, Wools-Kaloustian, Siika, Van
  Der~Laan, and Petersen]{tran2019double}
Linh Tran, Constantin Yiannoutsos, Kara Wools-Kaloustian, Abraham Siika, Mark
  Van Der~Laan, and Maya Petersen.
\newblock Double robust efficient estimators of longitudinal treatment effects:
  comparative performance in simulations and a case study.
\newblock \emph{The international journal of biostatistics}, 15\penalty0
  (2):\penalty0 20170054, 2019.

\bibitem[van~der Laan and Gruber(2011)]{van2011targeted}
Mark~J van~der Laan and Susan Gruber.
\newblock Targeted minimum loss based estimation of an intervention specific
  mean outcome.
\newblock 2011.

\bibitem[van~der Nest et~al.(2020)van~der Nest, Passos, Candel, and van
  Breukelen]{van2020overview}
Gavin van~der Nest, Val{\'e}ria~Lima Passos, Math~JJM Candel, and Gerard~JP van
  Breukelen.
\newblock An overview of mixture modelling for latent evolutions in
  longitudinal data: Modelling approaches, fit statistics and software.
\newblock \emph{Advances in Life Course Research}, 43:\penalty0 100323, 2020.

\bibitem[Vermunt(2010)]{vermunt2010latent}
Jeroen~K Vermunt.
\newblock Latent class modeling with covariates: Two improved three-step
  approaches.
\newblock \emph{Political analysis}, pages 450--469, 2010.

\end{thebibliography}
\bibliographystyle{plainnat}

\end{document}